\documentclass[doublecol]{epl2} 
\title{Harnessing individual nitrogen-vacancy centers with a compact and portable confocal microscope}
\shorttitle{Compact confocal microscope for NV Centers} %Insert here a short version of the title if it exceeds 70 characters

\author{I. Panadero\inst{1,2,3} \and JC. Guerra\inst{1} \and E. Caravaca \inst{1} \and F. J. Hidalgo \inst{2} \and  P. Acedo \inst{4} \and  C. de Dios \inst{4} \and E. Torrontegui \inst{2} }
\shortauthor{I. Panadero \etal}
\usepackage{amsmath}
\usepackage{physics}
\usepackage{ulem}
\usepackage{xcolor}

\institute{                    
  \inst{1} Arquimea Research Center - Camino las Mantecas s/n, 38320 San Cristobal de La Laguna, Spain\\
  \inst{2} Department of Physics - Universidad Carlos III de Madrid, Avda. de la Universidad 30, 28911 Legan\'es, Spain\\
  \inst{3} Department of Physical Chemistry, University of the Basque Country UPV/EHU, 48080 Bilbao, Spain\\
  \inst{4} Department of Electronic Technology, Universidad Carlos III de Madrid, Avda. de la Universidad 30, 28911 Legan\'es, Spain
}

\abstract{Recent advancements in quantum technology have highlighted the potential of nitrogen-vacancy (NV) centers in diamond. However, fully realizing this potential requires addressing challenges related to the size, complexity, and cost of current optical systems used for NV center manipulation. In this work, we present a compact and portable confocal setup specifically designed for the efficient detection and control of single NV centers. Our system facilitates optical initialization and readout of individual NV center photoluminescence signals, enabling coherent spin control and nanoscale-resolution magnetic field sensing.}

\begin{document}

\maketitle

\section{Introduction}

%Over the last two decades, quantum technologies (QT) have experienced significant progress, evolving from intricate experiments in quantum physics into a thriving multidisciplinary field of applied research \cite{acin2018quantum, dowling2003quantum}. We are witnessing advances in developing technologies that specifically manipulate individual quantum states and leverage distinctive properties, such as superposition and entanglement. This expanding field encompasses quantum communication, simulation, computation, and sensing. Various platforms, including trapped ions \cite{blatt2012quantum, haffner2008quantum}, neutral atoms \cite{henriet2020quantum}, quantum dots \cite{lim2015carbon}, and superconducting circuits \cite{devoret2013superconducting}, have emerged as promising candidates for QT applications. However, the sensitivity of quantum systems to environmental disturbances necessitates operation at cryogenic temperatures or inside vacuum chambers, contributing to the high cost of specialized equipment and personal \cite{hornibrook2015cryogenic}.
%

Nitrogen-vacancy (NV) centers in diamond have emerged as exceptionally promising candidates for the implementation of quantum technologies. These centers exhibit atom-like properties, characterized by long-lived spin quantum states and well-defined optical transitions, all within a robust solid-state device\cite{doherty2013nitrogen}. The electron spins of NV centers can be easily initialized, controlled, and read out at room temperature, simplifying the experimental setup compared to other quantum systems that require cryogenic temperatures. This provides practical advantages for the development and application of quantum technologies \cite{sewani2020coherent}. 

The adaptability of NV centers across different applications highlights their significance in advancing quantum technologies. NV centers serve as quantum repeaters in quantum communication \cite{jing2022quantum, nemoto2016photonic}, playing a crucial role in extending the range and improving the efficiency of quantum information transfer. Additionally, they function as quantum memories in the realm of quantum computing \cite{yang2011high}, facilitating the storage and retrieval of quantum information. Furthermore, these centers  have been proposed as quantum information processing units for quantum simulation \cite{cai2013large, ju2014nv}, contributing to the exploration and understanding of exotic many-body quantum systems. However, the domain where NV centers stands out is in quantum sensing being highly effective at detecting magnetic fields \cite{ho2021recent}, temperature \cite{wang2015high}, electric fields \cite{dolde2011electric}, and pressure \cite{ivady2014pressure}. All the mentioned applications rely on their high sensitivity,  large room-temperature quantum coherence up to milliseconds \cite{kennedy2003long}, and remarkable spatial, spectral and temporal resolution \cite{meinel2023high, radtke2019nanoscale, ninio2021high}. This characteristics enable the development of robust and calibration-free sensors. 

%The diamond host of NV centers adds significant advantages to their development. The exceptional properties of diamond \cite{childress2013diamond}, including strength, transparency, and excellent thermal conductivity, contribute favorably to the functionality of NV centers. Furthermore, the ability to fabricate photonic structures directly from the diamond crystal facilitates the creation of efficient optical interfaces \cite{jung2019spin}, and the compatibility of diamond with biological systems renders it highly suitable for quantum biosensing applications \cite{zhang2021toward}, particularly when utilizing nanodiamonds \cite{wang2023recent}. 

%Additionally, there is a demand for advanced fabrication techniques that can deterministically create NV centers with high spatial control \cite{chakraborty2019cvd}. Overcoming these obstacles not only propels the advancement of existing applications but also opens avenues for exploring new territories in quantum research and technology. 

%enhance the collection efficiencies for the emitted photons photons \cite{kaupp2016purcell}, a task that may be achieved through the implementation of cavities and optical nanostructures. Additionally, there is a demand for 

Despite their promise, NV center technologies face several development challenges \cite{radtke2019nanoscale, childress2013diamond}. One significant hurdle is the low photon emission rates of NV centers combined with the low collection efficiency of many setups \cite{kaupp2016purcell}. Additionally, achieving precise spatial control during NV center implantation remains problematic \cite{chakraborty2019cvd}. These challenges complicate the development of portable and accessible experimental instruments, hindering the scalability and broader application of this quantum platform \cite{xia2024design}.

NV center technology based on ensembles has been successfully miniaturized and integrated into compact platforms and systems \cite{pogorzelski2024compact, stegemann2023modular, misonou2020construction, zhao2023all, kim2019cmos, du2021high}. This progress not only reduces the physical footprint of the setups but also significantly cuts costs, making the technology more accessible to a wider range of industries and the general public. Some commercial systems have already been developed for educational activities \cite{qutools, ciqtek}. These advancements in size, compactness, and affordability have the potential to revolutionize applications in quantum technologies \cite{liu2018quantum}, as well as biomedical imaging \cite{balasubramanian2014nitrogen}. However, achieving effective single NV center detection remains challenging within miniaturized setups, as it still requires highly sensitive and efficient detection capabilities.  

The ability to detect single NV centers is essential for fully harnessing their quantum capabilities. In the field of quantum sensing and diamond-based technologies, the detection of single NV centers has been extensively investigated \cite{misonou2020construction, jelezko2001spectroscopy, nishimura2024investigations}. Techniques such as confocal microscopy and wide-field imaging have been used to explore NV centers, significantly advancing our understanding of quantum phenomena at the nanoscale \cite{scholten2021widefield, taylor2008high, kost2015resolving}. However, challenges remain with current experimental setups, mainly due to issues of size, complexity, and cost. Recent developments have introduced new processing techniques \cite{kudyshev2023machine} and optical components \cite{oeckinghaus2014compact} that improve sensitivity and efficiency in detecting single NV centers. Despite these advances, there is still a strong need for compact solutions that effectively balance performance and accessibility.

The growing interest for portable and cost-effective quantum sensing devices underscores the need for detecting individual NV centers with compact and portable setups. To meet this demand, we have developed a novel compact confocal setup of approximately 525 mm $\times$ 375 mm  $\times$ 300 mm (L $\times$ W $\times$ H) size. Contrary to standard portable devices \cite{pogorzelski2024compact, stegemann2023modular, misonou2020construction, zhao2023all, kim2019cmos, du2021high, qutools, ciqtek} this benchtop system allows the detection and control of individual NV centers incorporating recent technological advancements such as a designed optics that allows a size reduction maintaining the stability and resolution, ensuring practicality, cost-effectiveness, and ease of integration with other experimental components. 

By miniaturizing the traditionally bulky and complex confocal microscope setup, we gain the flexibility to conduct experiments in diverse environments outside the laboratory. This portability eliminates the need for specialized infrastructure and facilitates on-the-go measurements, promoting rapid data collection and analysis. Additionally, the compact design enhances accessibility, enabling institutions with limited resources to explore single NV center phenomena and applications.

Our microscope system can optically read out the photoluminescence (PL) signal from individual NV centers and coherently control their spin states, while addressing the need for efficiency and portability in quantum experiments. Beyond the single NV detection, the system accommodates also both nanodiamonds and ensembles, making it a versatile tool for many applications. Coherent control is achieved by precisely synchronizing laser and microwave (MW) pulses, allowing for magnetic field sensing with nanoscale resolution and detailed characterization of NV center spin properties. This instrument aims to overcome existing barriers and facilitate broader adoption and exploration of quantum phenomena at the single NV center level.

\section{The NV center in diamond}
\begin{figure}[t!]%
  \centering
\includegraphics[width= 1\linewidth, trim={0 0 2.5cm 0},clip]{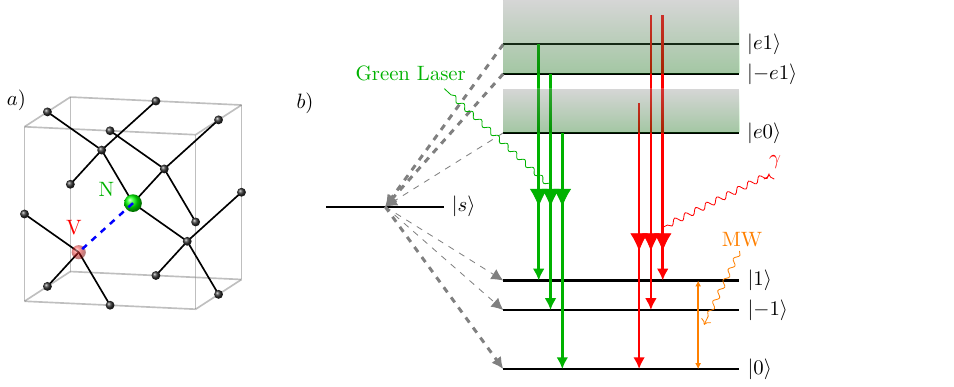} %
  \caption{(a) Atomic structure of the NV center in diamond, a substitutional nitrogen atom (green) adjacent to a vacant lattice site (transparent). The black dots represent the carbons atoms of the diamond lattice. (b) Representation of the levels of an NV center showing the system dynamics. The  $\ket{0}$ and $\ket{1}$ electronic ground states encode the qubit and are coupled with a MW field of Rabi frequency $\Omega$. Laser light excites the states $\ket{0}$ and $\ket{\pm 1}$ to the excited states $\ket{e0}$ and $\ket{\pm e1}$. The excited states decay to the ground state emitting a  $\sim 637-800$ photon. The states $\ket{\pm e1}$ decays to the singlet  $\ket{s}$ state with higher probability than the state  $\ket{0}$. The state $\ket{s}$ decays mainly to the $\ket{0}$ state.}%
  \label{fig:nv_scheme}%
\end{figure}

The NV center in diamond consists of a substitutional nitrogen atom adjacent to a vacant lattice site, as schematically shown in Fig. \ref{fig:nv_scheme}a. The NV center can exist in three charge states: the neutral NV$^0$, the positively charge NV$^+$, and the negatively charged NV$^-$ state. We are only interested in the NV$^-$ state because it is the only one which is magneto-optically active. Henceforth, we will refer to the NV$^-$ state simply as the NV center.

The NV center ground state corresponds to a spin triplet characterized by the $\{\ket{0},\ket{1},\ket{-1}\}$ states. Neglecting any nuclear or
spin-strain interactions, and considering the NV-axis oriented along the z-direction, parallel to an external applied magnetic field $B_0$, the static Hamiltonian of the ground state is given by:
\begin{equation}
    \hat H_0 = D \hat S_z^2 + \gamma_e B_0 \hat S_z,
    \label{eqn:hamiltonian}
\end{equation}
where $D = 2.87$ GHz is the zero field splitting, $\gamma_e = 28$~GHz/T is the electron gyromagnetic ratio, and $\hat S_{x, y, z}$ are the spin-1 matrices. The second term in Eq.~\eqref{eqn:hamiltonian} accounts for the Zeeman splitting of the states $\ket{\pm 1}$. Coherent control of the internal levels is achieved by means of a resonant MW signal generated by an antenna with either the $\{\ket{0},\ket{1}\}$ or $\{\ket{0},\ket{-1}\}$ transitions depending on the selected qubit encoding. Such oscillating field is given by the time-dependent Hamiltonian:
\begin{equation}
    \hat H_1 = \Omega \cos(\omega t) \hat S_x,
\end{equation}
with $\Omega$ and $\omega$ the amplitude and frequency of the driving field.

In order to describe the spin initialization and readout mechanisms \cite{panadero2024photon}, we must consider the optically excited triplet $\{\ket{e0},\ket{e1},\ket{-e1}\}$ states
and the metastable singlet $\ket{s}$ state, depicted in Fig.\ref{fig:nv_scheme}(b), along the already introduced ground states. Non-resonant laser light is used to couple the ground states to the phononic band of the excited-state manifold. These vibrational states quickly decay into the corresponding spin-conserving excited state, destroying the optical coherence between the ground and excited manifolds \cite{nizovtsev2001}. The excited states relax either directly through a spin-conserving transition emitting red fluorescence wavelength of approximately  637–800 nm, or through a non-spin-conserving channel through the metastable spin-singlet $\ket{s}$ states responsible for the initialization and readout of the NV. Optical readout is possible due to a decay imbalance, where the $\ket{\pm e1}$ states decay faster than the $\ket{e0}$ state through the non-radiative transitions involving the $\ket{s}$ state. As result, the $\ket{0}$ state appears brighter than the $\ket{\pm 1}$ states when excited, because a greater proportion of the population relaxes back to the ground state through radiative transitions in that case. Furthermore, optical polarization occurs because the  $\ket{s}$ state decays with higher rate to the $\ket{0}$ state than to the $\ket{\pm 1}$ states, allowing for qubit initialization.

\section{Materials and Methods}

Harnessing single NV centers relies heavily on optical methods, therefore the optical components of the confocal microscope are crucial. The Zeeman splitting of the NV center ground state is detected through changes in the spin-dependent PL signal, indicating that the readout process is optical in nature. As a result, the ability to detect a decrease in PL intensity at resonance is directly related to the sensitivity of magnetic field sensing using NV centers and the overall performance of the device. 

\begin{figure}[t!]%
  \centering
\includegraphics[width= 1\linewidth]{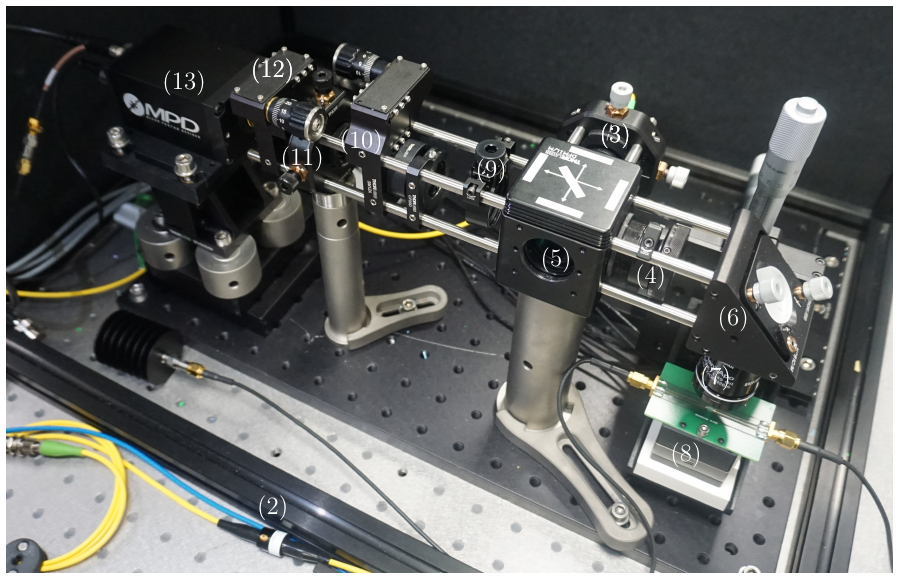} %
  \caption{Picure of the optical elements of the confocal setups, labeled as they appeared in the main text.}%
  \label{fig:confocal_picture}%
\end{figure}

\begin{figure}[ht]%
  \centering
\includegraphics[width= 1\linewidth]{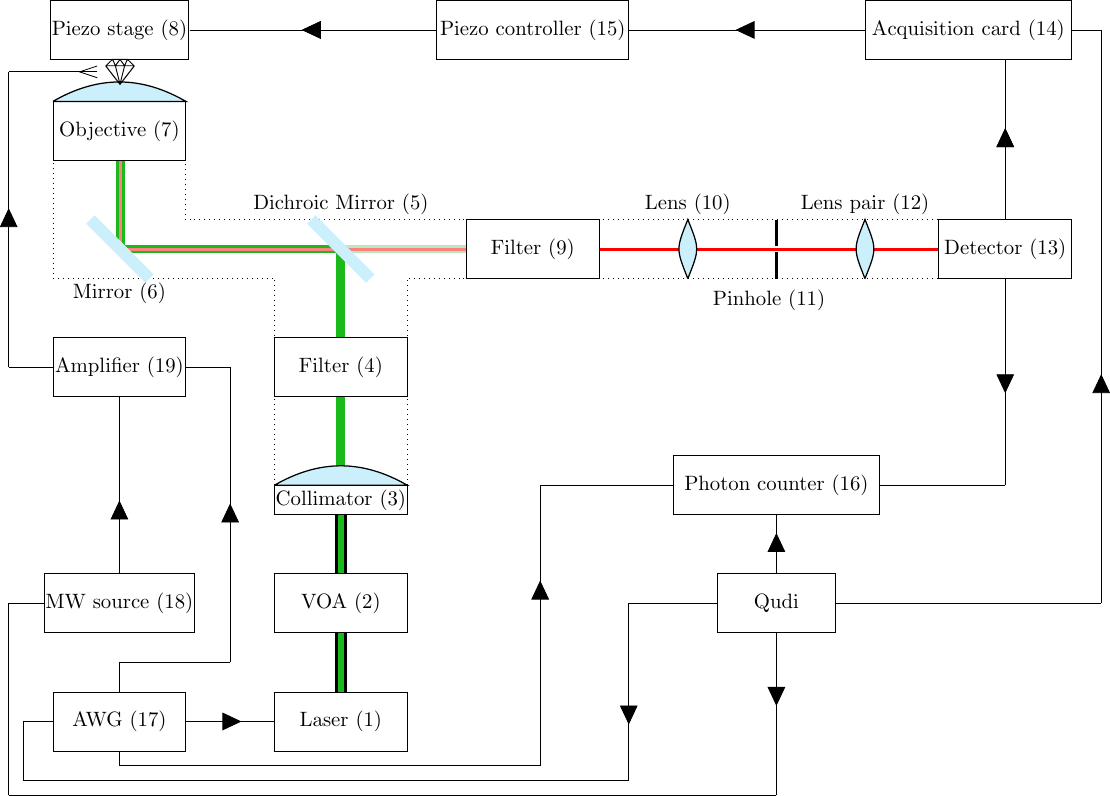} %
  \caption{Schematics of the main components of the compact confocal microscope and the devices used.}%
  \label{fig:SetupSchematics}%
\end{figure}

In Fig. \ref{fig:confocal_picture} we provide a picture of the developed compact confocal system, while Fig. \ref{fig:SetupSchematics} depicts a schematic overview of its components. The optical elements are mounted into a cage system which provides a good stability and alignment of the optical elements. To excite the NV center, a green laser (1 - Toptica iBEAM SMART PT 515) emitting collimated light at a wavelength of $515$ nm is employed. The laser power can be finely adjusted using a fiber variable optical attenuator (2 - Thorlabs V450A).

The laser beam is guided through a single-mode fiber and collimated (3 - Thorlabs F810APC-543). Subsequently, the beam passes through a $550$ nm low-pass filter (4 - Thorlabs FESH0550), effectively eliminating second harmonics originating from either the laser or the fiber. Next, the beam is directed towards a dichroic mirror (5 - Thorlabs DMLP550R), which then reflects the beam onto another mirror (6 -Thorlabs BB1-E02) that, in turn, redirects the light towards the microscope objective (MO) aperture (7 - Olympus UPLXAPO 100X) immersed in oil to reduce photon scattering. 

Maintaining a beam size equal to the MO aperture diameter ensures that the focused spot size at the sample closely matches the diffraction-limited resolution.  The MO is crucial in a confocal microscope because it determines the efficiency of photon collection and spatial resolution. Our MO has a numerical aperture of 1.45, a magnification of 100, and a working distance of $135$ $\mu$m. 

The detection path begins in the MO focal plane where light is collected and collimated. The detection and laser path are split at the dichroic mirror, which transmits the longer wavelength PL light and blocks the reflected excitation light. Another 650 nm longpass filter (9 - Thorlabs FELH0650) is placed in the optical path  to enhance the signal-to-noise ratio by further isolating the PL signal from the  residual reflected excitation light and the PL of the NV$^-$ from that of the NV$^0$. The collected PL is then focused with a 50 mm lens (10 - Thorlabs AL2550M-A) into a pinhole (11 - Thorlabs P20HK) of $20$ $\mu$m. The pinhole enhances the axial and lateral resolution by allowing only the focused light to contribute to the image, reducing background noise.

The light exiting the pinhole is focused by a lens pair (12 - Thorlabs MAP051919-A) into the photon detector, a single-photon avalanche diode (SPAD) with single-photon sensitivity (13 -MPD PDM-050-CTB). The SPAD provides high sensitivity for detecting low photon rates and enables nanosecond resolution for the PL signal readout.  When a photon arrives at the SPAD, it outputs a electrical pulse delivered into the Time-Correlated Single Photon Counting (TCSPC) card (16 - PicoQuant TimeHarp 260) for time-tagging analysis. Alternatively, for simple PL experiments, the signal can be delivered to a pulse counting data acquisition card (14 - NI USB-6343 (BNC)). For applications where the spectral composition of the optical signal is of interest, the SPAD can be substituted by a spectrometer.

For precise alignment, the pinhole is affixed to an $xyz$-translation stage incorporated within the cage system, allowing meticulous adjustments, with the $z$ axis aligned parallel to the beam. The lenses are also mounted onto a $z$ stage for accurate positioning. Lastly, the SPAD is situated on an $xyz$ stage, ensuring stable positioning during experimentation. 

To scan the sample beneath the MO, we secure it onto a nanopositioner featuring a piezo-based stage with a 100~$\mu$m travel range, enabling nanometer-resolution motion along the $xyz$ axes (8 - Newport NPXYZ100SG). The position of the stage is controlled using three analogue signals ranging from 0-10 V, one for each axis, with 0 V corresponding to the the most retracted position on that axis (0 $\mu$m) and 10 V to full extension (100 $\mu$m) (15 - Newport NPC3SG). This piezo controller is then connected to the acquisition card to manage the confocal scans. This nanopositioner is then integrated into a coarse stage, facilitating the mounting and positioning of the sample. 

The MW excitation signal, crucial for controlling the spin of the NV center,  is achieved using a 20 $\mu m$ cable securely affixed to the diamond surface and terminated with at 50 $\Omega$ load. Pulsed sequences are generated using an Arbitrary Waveform Generator (AWG) (17 - Keysight / M8190A), while a simpler signal generator manages the continuous sequences (18 - Windfreak SynthHD). A $45$ dBm amplifier allows to boost the strength of the MW signals (19 - MiniCircuits ZHL - 16W-43-S+). 

%When fully assembled, the instrument fits within a black box measuring 525 mm x 375 mm x 300 mm, with the exception of the control electronics, which are accommodated on a separate rack. This setup is designed to provide optical isolation from external light sources and to confine the working environment to prevent open beam exposure.

Qudi, a specialized software \cite{binder2017qudi}, orchestrates the scanning process, manages PL signal acquisition and processing, and controls both laser and MW pulses.  It delineates a structured framework by segregating its functionalities into distinct layers: hardware abstraction, experiment logic, and user interface. At its core, Qudi offers an extensive array of features including a user-friendly graphical interface, real-time data visualization, robust configuration management, and comprehensive data recording capabilities.

 The confocal microscope is mounted on a breadboard with dimensions of 450 mm $\times$ 300 mm (W $\times$ H), allowing it to be transported without dismantling or causing optical misalignment. Including the breadboard, the benchtop system has a relatively light weight of approximately 15 kg. Precision in optomechanical design ensures that the confocal setup maintains its alignment over time. The breadboard can be fixed to an optical table for stability. When fully assembled, the instrument fits within a black optical enclosure measuring 525 mm $\times$ 375 mm  $\times$ 300 mm (L $\times$ W $\times$ H), with the exception of the control electronics, which are accommodated on a separate rack. This optical enclosure is designed to provide optical isolation from external light sources and to confine the working environment to prevent open beam exposure.

The compactness and portability of this confocal allows for easy transportation between different laboratory setups or field locations, making it versatile for various research environments. The reduced footprint ensures that the system can be integrated into crowded laboratory spaces without requiring significant rearrangement of existing equipment. Furthermore, the portable nature of the setup enables rapid deployment and installation, which is beneficial for time-sensitive experiments and collaborations across different research facilities. The robust design minimizes the risk of misalignment during transport, ensuring that the system is ready for immediate use upon arrival. Overall, the compact and portable nature of this confocal microscope enhances its utility and flexibility in diverse scientific and industrial applications.

\section{Results}

\begin{figure}[t!]%
  \centering
\includegraphics[width= 0.85\linewidth]{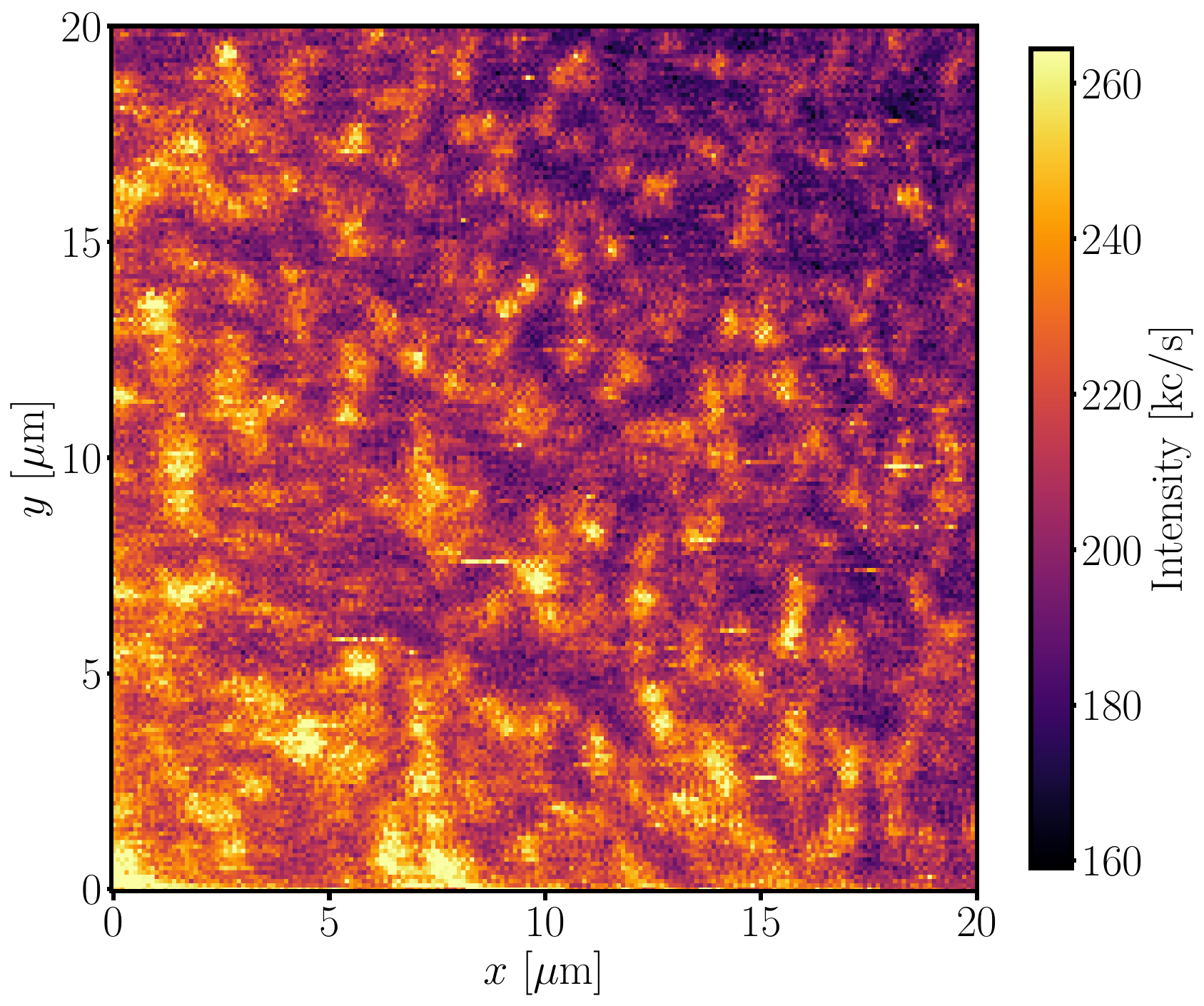} %
  \caption{Confocal PL scan of the sample of area  20 $\mu$m $\times$ 20 $\mu$m with a 10 nm resolution. Bright spots correspond to single NV centers.}%
  \label{fig:single_NV_confocal}%
\end{figure} 

The sample used to detect single NV centers and to benchmark the setup is an implanted Single Crystal Electronic (ELSC) grade diamond from ElementSix, which is a high-purity chemical vapor deposition (CVD) substrate material. The implantation energy is set at $5$ eV, positioning the NV centers $10$ nm below the surface, and are implanted with a dose of $10^9$ $^{15}\text{N}^+/\text{cm}^3$. In Fig. \ref{fig:single_NV_confocal} we observe a confocal 20 $\mu$m $\times$ 20 $\mu$m scan, with a 10 nm resolution, where the bright dots correspond to single NV centers. 

\begin{figure}[ht]%
  \centering
\includegraphics[width= 1\linewidth]{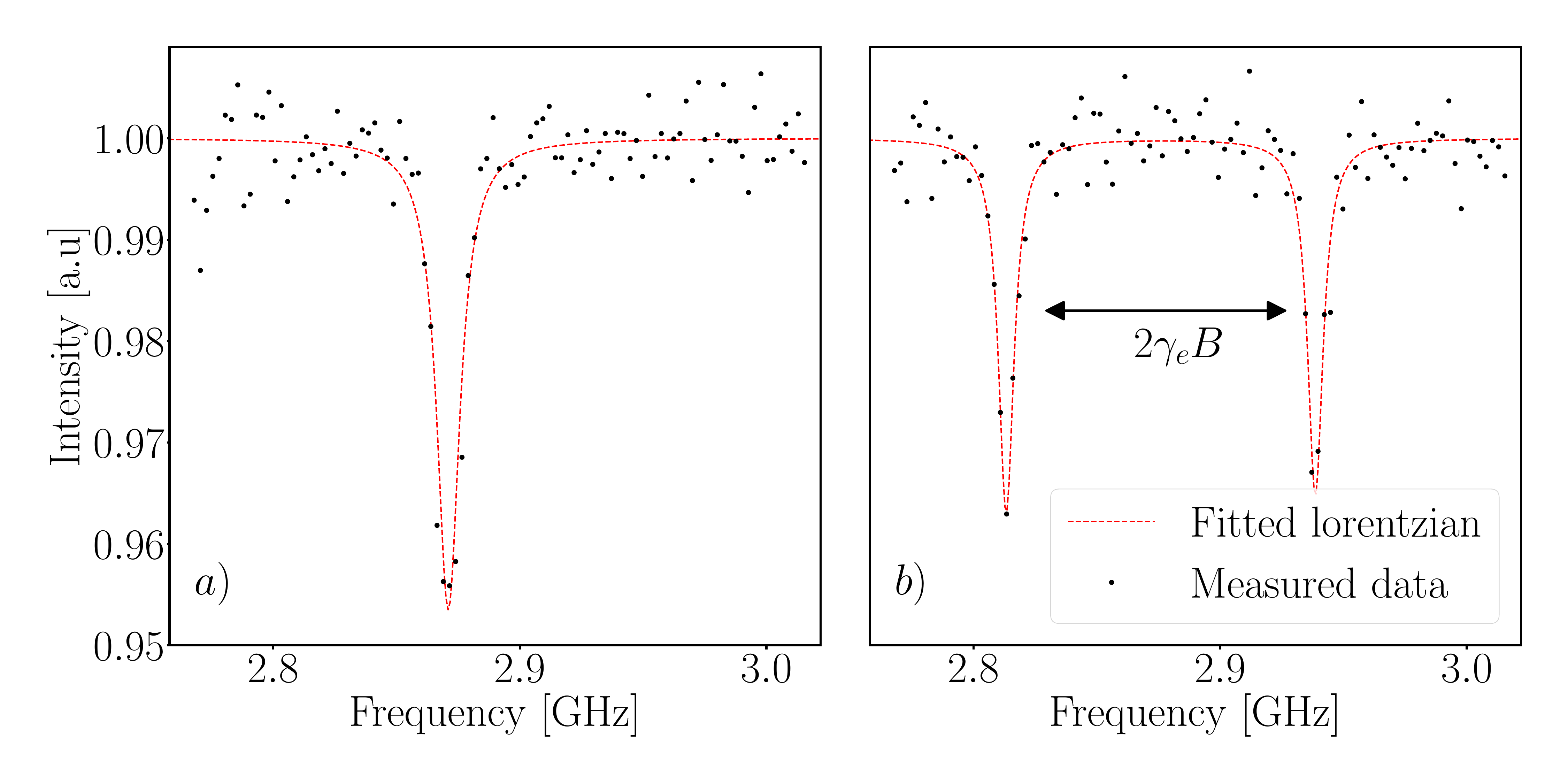} %
  \caption{a) ODMR spectrum at zero magnetic field for a single NV center. b) ODMR spectrum for an arbitrary magnetic field applied to the NV center. The central resonance is split into two resonances due to the Zeeman splitting of the $\ket{\pm 1}$ states. }%
  \label{fig:SNV_ODMR}%
\end{figure}

To measure the spin transition spectrum of the single NV center, the laser continuously excites the NV center while we swept the frequency of the MW rotating the spin of the ground state. When the MW frequency equals the $\ket{0} \leftrightarrow \ket{\pm 1}$ transition frequency, population is transferred from the bright state $\ket{0}$ to the darker states $\ket{\pm 1}$, measuring a drop in the PL. This procedure is referred to as optically detected magnetic resonance (ODMR). Figure \ref{fig:SNV_ODMR}(a) shows the  ODMR spectrum at zero magnetic field.  The resonance  Full Width at Half Maximum (FWHM) is 11 MHz and the contrast is $4$\%, defined as one minus the ratio between the signal intensity at resonance and non-resonant conditions.

Introducing an external magnetic field produces the splitting of the central resonance due to the Zeeman interaction, see Eq.~\eqref{eqn:hamiltonian}. We present a ODMR spectra recorded under a magnetic field applied by positioning a permanent magnet near the diamond in Fig.~\ref{fig:SNV_ODMR}(b). The Zeeman energy splitting of the $\ket{\pm 1}$ states depends on the orientation of the magnetic field relative to the NV center axis. The component of the magnetic field along the NV center axis, denoted as $B_z$, results in a measurable Zeeman splitting, which can be observed and quantified from the spectra:
\begin{equation}
    \nu_+-\nu_- = 2\gamma_e B_z
\end{equation}
where $\nu_+$ and $\nu_-$ are the central frequencies of the two resonances, corresponding to the transitions $\ket{0} \leftrightarrow \ket{+1}$ and $\ket{0} \leftrightarrow \ket{-1}$ respectively. For the ODMR measurement in Fig. \ref{fig:SNV_ODMR}(b), the value of the magnetic field projection is $B_z = 356 \pm 1$ $\mu$T.  

In the previous experiment, we applied the MW and laser continuously, but pulsing the laser and MW provides higher optical contrast and better control of the spin of the NV center. Most experimental sequences start with a laser pulse to polarize the NV center into the $\ket{0}$ state. The spin state is then controlled through a sequence of MW pulses, after which it is read out again with a laser pulse. In many scenarios the initialization and readout pulse are combined. The MW pulses usually consist of sequences of $\pi$ and $\frac{\pi}{2}$ pulses in resonance with a selected transition. For example, a $\pi$-pulse transforms the state $\ket{0}$ into the state $\ket{1}$ and vice-verse, while a $\pi$/2-pulse brings the state $\ket{0}$ into the superposition $(\ket{0} + \ket{1})/\sqrt{2}$. The required duration of these pulses is measured through Rabi oscillations.

After successfully performing continuous wave ODMR on a single NV center, the next step is to measure Rabi oscillations -i.e. a periodic exchange of population between the $\ket{0}$ and $\ket{1}$ states under an external periodic driving MW field, indicating the coherent control of the enconded qubit, a crucial step in proving the viability of any quantum system.

\begin{figure}[t!]%
  \centering
\includegraphics[width= 1\linewidth]{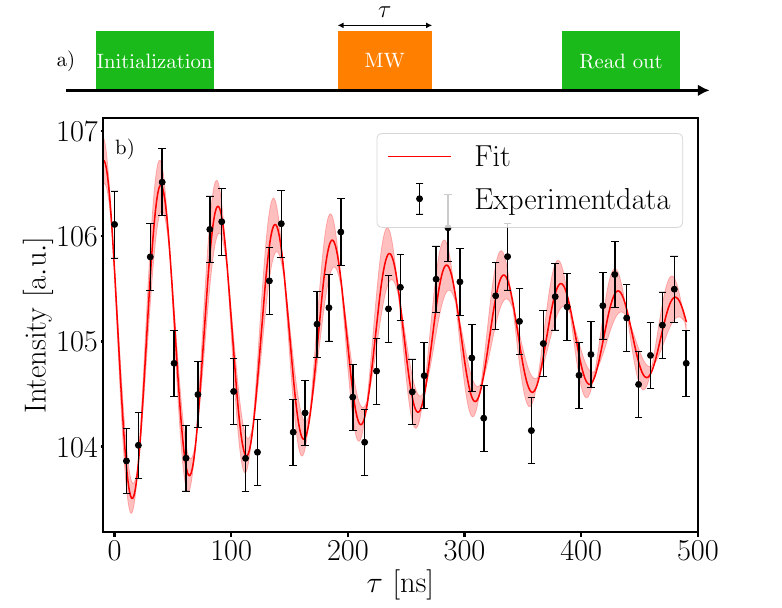} %
  \caption{a) Pulse sequence to perform Rabi oscillations. The length of the laser pulse is 2 $\mu$s and the length of the MW pulse is varied. b) Rabi oscillations for a single NV center in the spot 2 with MW frequency set to $\omega$ = 2.87 GHz. We measured for 1 minute and obtained $\Omega = 20.4 \pm 0.1$ MHz and $T_2^{*} = 320 \pm 60$ ns}%
  \label{fig:rabi_spot2}%
\end{figure}

To observe Rabi oscillations, we set the driving MW frequency to be on resonance with the ODMR transition frequency and vary the MW pulse length. An initial laser pulse of 3 $\mu$s is used to initialize the NV center into the $\ket{0}$ state. A subsequent MW pulse, of varying length $\tau$, then rotates the spin. Finally, a laser pulse of 3 $\mu$s reads out the NV center state. The schematic of the implemented pulse sequence for this experiment is shown in Fig. \ref{fig:rabi_spot2}(a). We observe oscillations in the PL signal as a function of MW pulse length, as shown in Fig. \ref{fig:rabi_spot2}(b), indicating coherent population transfer. The oscillations can be fitted to an exponentially decaying sinusoidal function
\begin{equation}
    S_{\text{Rabi}} = a\cos{(2\pi \Omega \tau + \phi)}\exp{-\tau/T_2^{*}} + c,
\end{equation}
where $a$ is the Rabi oscillation amplitude, $c$ is an offset, $\Omega$ is the Rabi frequency, $\phi$ is a phase, and $T_2^{*}$ is the driven coherence time of the spin. The Rabi frequency $\Omega$ characterizes the  coupling strength between the driving MW field and the NV center. The driven coherence time $T_2^{*}$ describes how long the $\ket{0} \leftrightarrow \ket{\pm 1}$ transition can be driven before the system dephases into a mixed-state. For the observed Rabi oscillations we determine the Rabi frequency $\Omega = 20.4 \pm 0.1$ MHz. From the Rabi frecuency value we calibrate the precise duration of the $\pi$/2-pulse and $\pi$-pulse  as $\tau_{\pi/2} = 12.25 \pm 0.05$ ns and $\tau_\pi = 24.5 \pm 0.1$  ns\cite{zhang2020diamond}. We also measured an intrinsic cpherence time $T_2^{*} = 320 \pm 60$ ns, consistent with the $T_2^*$ value of a shallow NV center \cite{irber2021robust, wang2016coherence}.  

In an NV center, the intrinsic dephasing time $T_2^*$ is influenced by magnetic noise from the enviroment, such as fluctuating spins interacting with the NV center spin, on a timescale of a few $\mu$s. These external spins include nuclear spins on the diamond lattice or unpaired electrons from dangling bonds at the surface. Consequently, $T_2^*$ depends on the NV center environment, such as the nitrogen concentration in the diamond and the distance between the NV center and the surface. While bulk NV centers can have coherence times up to a couple of milliseconds, shallow NV centers often exhibit dephasing times in the range of 1–100 $\mu$s \cite{tetienne2018spin}. Similar to nuclear magnetic resonance (NMR), the effects of magnetic noise can be further suppressed using specific pulse sequences, such as the Hahn echo, which will be implemented later in this work, or more sophisticated dynamical decoupling techniques \cite{wang2012comparison, pham2012enhanced}.

\begin{figure}[t!]%
  \centering
\includegraphics[width= 1\linewidth]{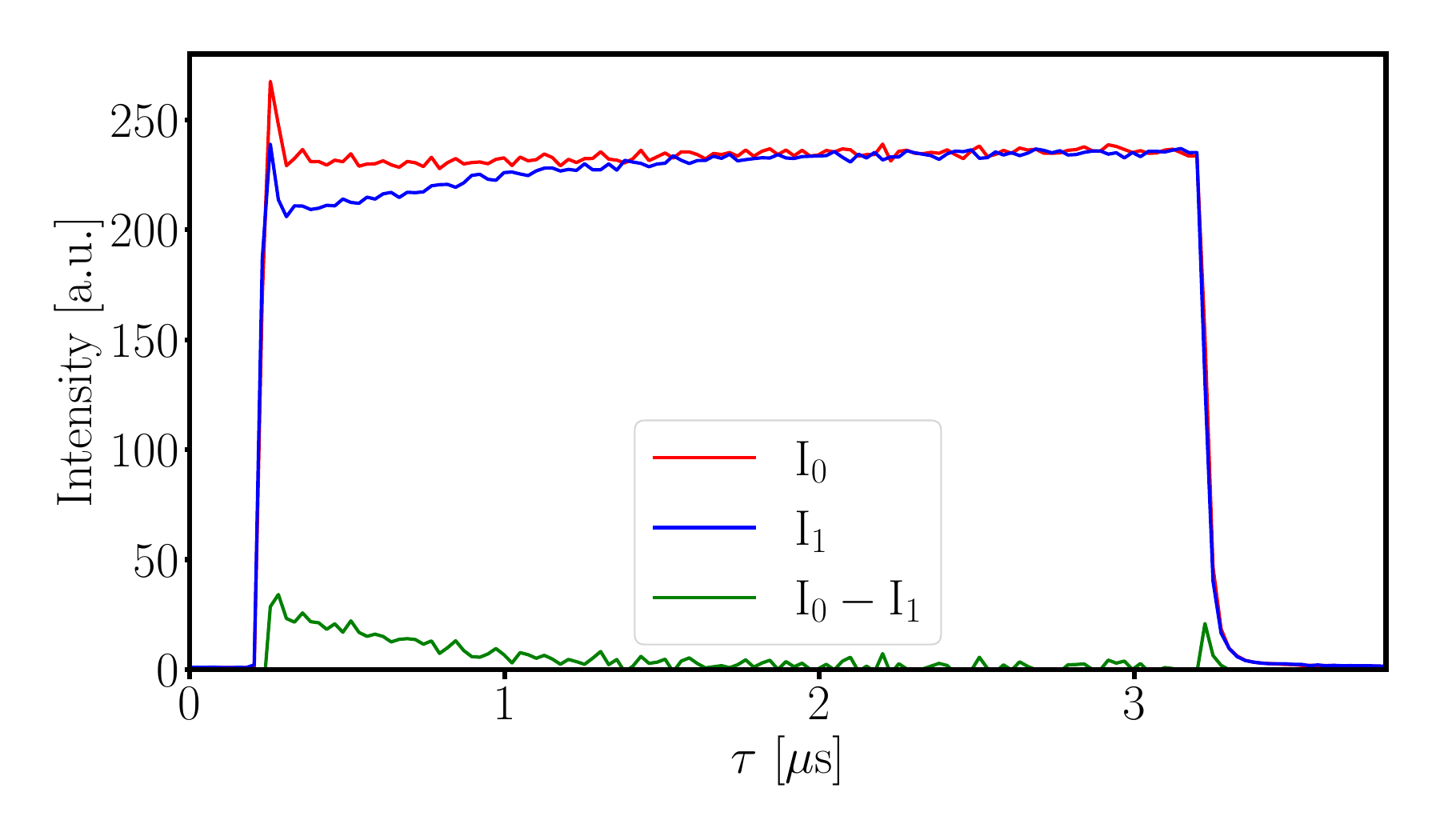} %
  \caption{Examples of time-dependent PL, denoted as $I_0$ and $I_1$, corresponding to pure initial spin states $\ket{0}$ and $\ket{1}$ for a single NV center. The PL is more intense when the system is initialized in the $\ket{0}$ state. After a time span of $1.5$ $\mu$s, both initializations reach the same steady state, causing their intensities to become equal. The contrast in the PL $|I_0-I_1|$ of the spin states is demonstrated in the lower half of the plot.}%
  \label{fig:intensity_I0I1}%
\end{figure}

The PL intensity provides insights into the state of the NV center. In Fig. \ref{fig:intensity_I0I1}, we present the time-dependent PL after an initialization and readout single laser pulse, denoted as $I_0$ (and $I_1$), for an initial $\ket{0}$ (or $\ket{1}$) state. To initialize the NV center into the $\ket{0}$ state, a 3 $\mu$s laser pulse is applied while initializaion in the $\ket{\pm 1}$ excited state requires the application of a subsequent MW $\pi$-pulse. 
%\IPM{The optical transitions involved in the} initialization process do not result in a pure $\ket{0}$ or $\ket{\pm 1}$ state. \ETM{Explicar errores de por qué no es ideal.} \IPM{The fluorescence intensity at the start of the pulse depends on the initial spin state of the NV centre \cite{panadero2024photon, doherty2013nitrogen}} 
The contrast in the readout of the NV center spin state corresponds to the small difference between the PL time traces $|I_0-I_1|$. We are able to detect this small difference thanks to the single photon sensitivity of our detector. Optical readout is performed by comparing the integrated fluorescence of the spin state being read out. Note that after approximately 1.5 $\mu$s, see Fig. \ref{fig:intensity_I0I1}, the NV center reaches a steady state with the same PL intensity, depolarizing due to optical excitation and resulting in a loss of information about the initial state. %\ETM{se puede explicar por que este numero?} \IP{Es por lo que se ve en la figura, en ese tiempo las dos trazas se igualan, pero para otra potencia de laser podría ser más o menos}.  \ETM{describir inset y decir que pese a la pequeña diferencia entre $I_0$ e $I_1$, uno puede inferir el estado del qubit [ref paper nuestro], comentar también cual es la diferencia mínimia en $I_0$ e $I_1$ detectable.} \IP{¿Qué te parece cómo está ahora? no entiendo qué quieres decir con ´diferencia mínima detectable', ya que eso dependerá del detector.}

\begin{figure}[t!]%
  \centering
  \includegraphics[width= 1\linewidth]{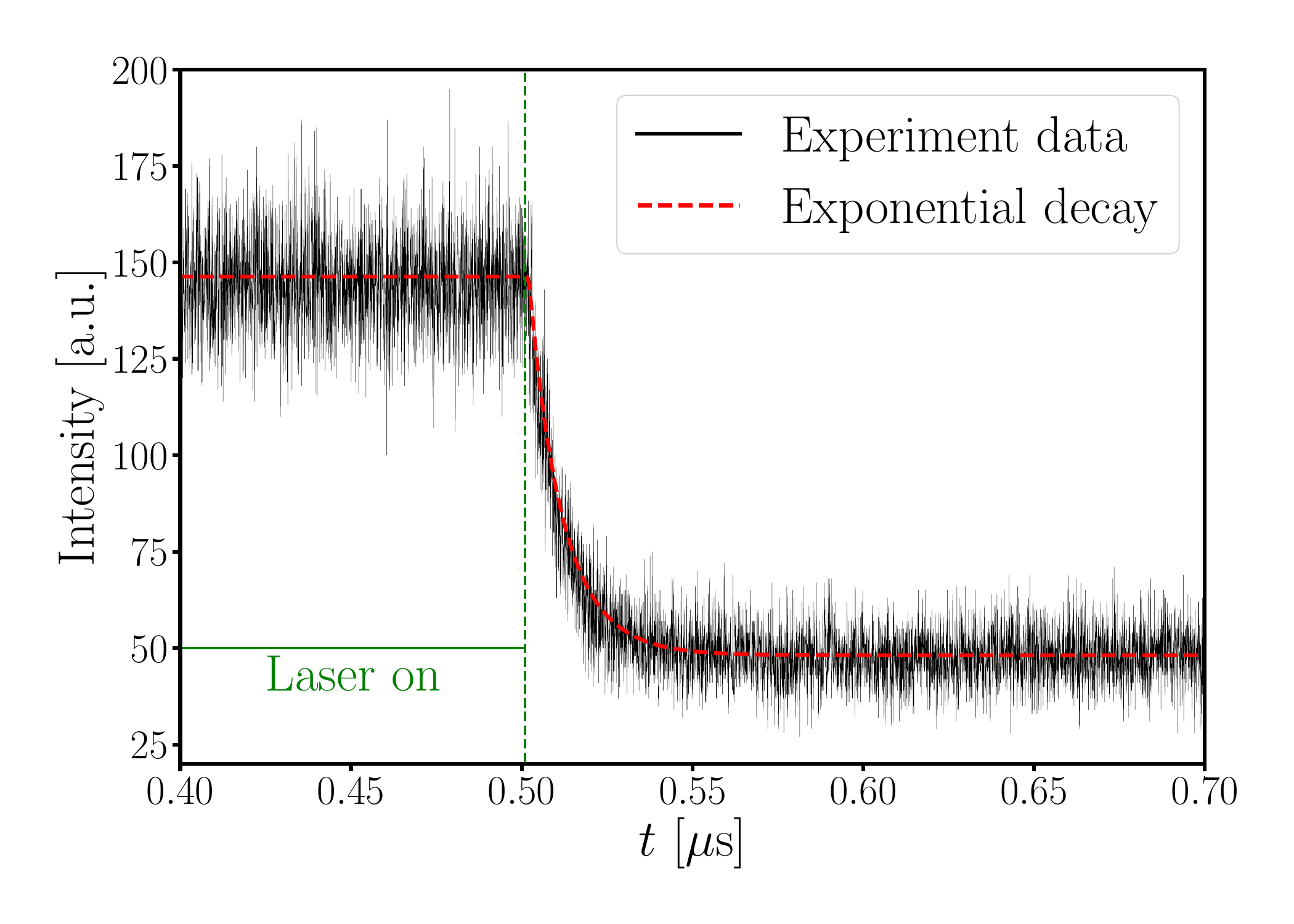} %
  \caption{Fluorescence lifetime measurement for a single NV center. Once the excitation is switched of the fluorescence rapidly decays with a lifetime of $12 \pm 2$ ns.}%
  \label{fig:nv_lifetime_arc}%
\end{figure}

After turning off the laser excitation, the population rapidly decays with a lifetime of $\sim 20$ ns from the excited states to the $\ket{0}$ ground state \cite{storteboom2015lifetime}, leading to a swift reduction in PL and the initialization of the NV center into the $\ket{0}$ state. This decay process is crucial for characterizing the transition rates between different energy levels of the NV center. From Fig. \ref{fig:nv_lifetime_arc} we measure a PL lifetime time of $12 \pm 2$ ns, consistent with other experimental results, where the NV center fluorescence lifetime typically ranges from 10 to 30 ns \cite{tisler2009fluorescence}.

\begin{figure}[t!]%
  \centering
  \includegraphics[width= 0.9\linewidth]{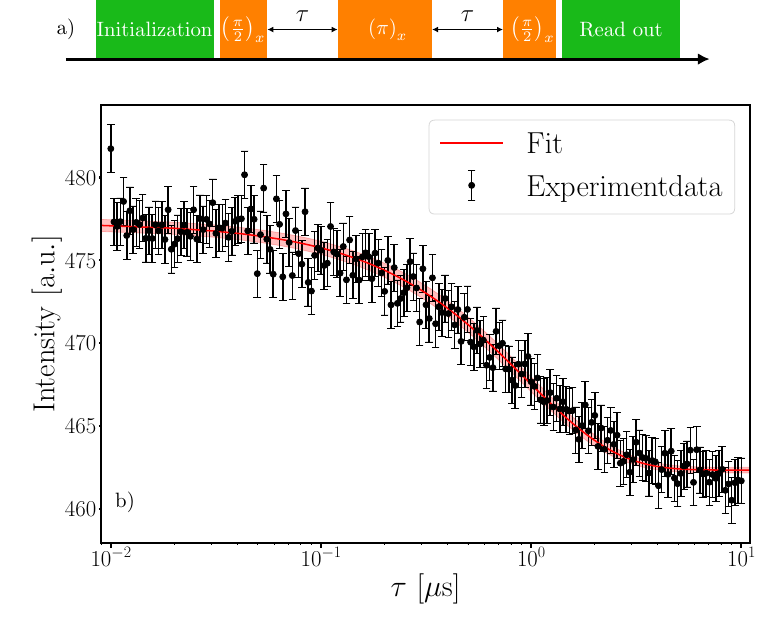} %
  \caption{a) Pulse sequence to enlarge the NV center dephaing time. The
    length of the laser pulse is 2 $\mu$s while the interspacing $\tau$ -i.e. free precesion time, is increased. b) Hahn Eco PL decay for a single NV center leading to $T_2
 = 940 \pm 40$ ns}%
  \label{fig:hanh_eco_spot4}%
\end{figure}

In the final measurement, we will assess the coherence time $T_2$ of the NV center using a Hahn Eco experiment. A Hahn echo is a dynamical decoupling sequence commonly used in the context of NMR and electron spin resonance (ESR) \cite{hahn801950} to protect the spin dynamics and extend its bare coherence time $T_2^*$. Unlike quantum systems operating at lower temperatures, room-temperature systems face significant thermal noise. Therefore, exploring the coherence times in room-temperature systems is particularly relevant. In Fig. \ref{fig:hanh_eco_spot4}(a), we present a schematic of the Hahn echo sequence. Initially,  the spin of the NV center is prepared in the $\ket{0}$ state. Subsequently, a  $\pi/2$-pulse is delivered to induce a 90-degree rotation of the spin. Following this, the spin evolves freely during a precession time $\tau$.  A subsequent  $\pi$-pulse  flips the spin, and after another precession period $\tau$, a second $\pi/2$-pulse is applied before the spin state is read out. The second free precession period helps to cancel out any uncontrolled phase. Experimental data shown in Fig. \ref{fig:hanh_eco_spot4}(b) is obtained by varying $\tau$ until the phase relation is randomized and the spin signal saturates, indicating a completely mixed state. Fitting the data to an exponential decay yields to an extended coherence time of $T_2 = 940 \pm 40$ ns. This enhancement compared to the measured intrinsic $T_2^*$ is attributed to the spin refocusing achieved by the Hahn echo sequence, which effectively reduces dephasing caused by low-frequency noise.

\section{Outlook and conclusions}

We have developed a compact confocal experimental setup optimized for single NV center detection and coherent spin control. This system enables precise measurements of magnetic resonance and delivery of pulse sequences. Its versatility makes it a valuable tool for various quantum technology applications, including the characterization of spin defects in solids, nanoscale NMR, hyperpolarization, bio-sensing, quantum sensing, and quantum information processing. Although further efforts are needed to miniaturize and enhance the accessibility of the system, the integration of a single NV center detector into a portable, compact confocal microscope significantly streamlines experimental workflows, accelerates research progress, and broadens the potential of NV center technology.

\acknowledgments
We thank Fedor Jelezko, Boris Naydenov, and Maabur Sow for fruitful discussions and the diamond sample implantation. We acknowledge financial support form the Spanish Government via the project PID2021-126694NA-C22 (MCIU/AEI/FEDER, EU), and project TSI-069100-2023-8 (Perte Chip-NextGenerationEU), from Horizon Europe project QCircle No 101059999 (Teaming for Excellence), and from Arquimea Resserch Center. E.T. acknowledge the Ram\'on y Cajal (RYC2020-030060-I) research fellowship.

\bibliography{quantum_systems}

\begin{thebibliography}{10}
\expandafter\ifx\csname url\endcsname\relax\def\url#1{\texttt{#1}}\fi

\bibitem{doherty2013nitrogen}
\Name{Doherty M.~W., Manson N.~B., Delaney P., Jelezko F., Wrachtrup J. \and Hollenberg L.~C.} \REVIEW{Physics Reports}{528}{2013}{1}.

\bibitem{sewani2020coherent}
\Name{Sewani V.~K., Vallabhapurapu H.~H., Yang Y., Firgau H.~R., Adambukulam C., Johnson B.~C., Pla J.~J. \and Laucht A.} \REVIEW{American Journal of Physics}{88}{2020}{1156}.

\bibitem{jing2022quantum}
\Name{Jing Y. \and Razavi M.} \REVIEW{Physical Review Applied}{18}{2022}{024041}.

\bibitem{nemoto2016photonic}
\Name{Nemoto K., Trupke M., Devitt S.~J., Scharfenberger B., Buczak K., Schmiedmayer J. \and Munro W.~J.} \REVIEW{Scientific reports}{6}{2016}{26284}.

\bibitem{yang2011high}
\Name{Yang W., Yin Z., Hu Y., Feng M. \and Du J.} \REVIEW{Physical Review A}{84}{2011}{010301}.

\bibitem{cai2013large}
\Name{Cai J., Retzker A., Jelezko F. \and Plenio M.~B.} \REVIEW{Nature Physics}{9}{2013}{168}.

\bibitem{ju2014nv}
\Name{Ju C., Lei C., Xu X., Culcer D., Zhang Z. \and Du J.} \REVIEW{Physical Review B}{89}{2014}{045432}.

\bibitem{ho2021recent}
\Name{Ho K.~O., Wong K.~C., Leung M.~Y., Pang Y.~Y., Leung W.~K., Yip K.~Y., Zhang W., Xie J., Goh S.~K. \and Yang S.} \REVIEW{Journal of Applied Physics}{129}{2021}{241101}.

\bibitem{wang2015high}
\Name{Wang J., Feng F., Zhang J., Chen J., Zheng Z., Guo L., Zhang W., Song X., Guo G., Fan L. \etal} \REVIEW{Physical Review B}{91}{2015}{155404}.

\bibitem{dolde2011electric}
\Name{Dolde F., Fedder H., Doherty M.~W., N{\"o}bauer T., Rempp F., Balasubramanian G., Wolf T., Reinhard F., Hollenberg L.~C., Jelezko F. \etal} \REVIEW{Nature Physics}{7}{2011}{459}.

\bibitem{ivady2014pressure}
\Name{Iv{\'a}dy V., Simon T., Maze J.~R., Abrikosov I. \and Gali A.} \REVIEW{Physical Review B}{90}{2014}{235205}.

\bibitem{kennedy2003long}
\Name{Kennedy T., Colton J., Butler J., Linares R. \and Doering P.} \REVIEW{Applied Physics Letters}{83}{2003}{4190}.

\bibitem{meinel2023high}
\Name{Meinel J., Kwon M., Maier R., Dasari D., Sumiya H., Onoda S., Isoya J., Vorobyov V. \and Wrachtrup J.} \REVIEW{Communications Physics}{6}{2023}{302}.

\bibitem{radtke2019nanoscale}
\Name{Radtke M., Bernardi E., Slablab A., Nelz R. \and Neu E.} \REVIEW{Nano Futures}{3}{2019}{042004}.

\bibitem{ninio2021high}
\Name{Ninio Y., Waiskopf N., Meirzada I., Romach Y., Haim G., Yochelis S., Banin U. \and Bar-Gill N.} \REVIEW{Acs Photonics}{8}{2021}{1917}.

\bibitem{childress2013diamond}
\Name{Childress L. \and Hanson R.} \REVIEW{MRS bulletin}{38}{2013}{134}.

\bibitem{kaupp2016purcell}
\Name{Kaupp H., H{\"u}mmer T., Mader M., Schlederer B., Benedikter J., Haeusser P., Chang H.-C., Fedder H., H{\"a}nsch T.~W. \and Hunger D.} \REVIEW{Physical Review Applied}{6}{2016}{054010}.

\bibitem{chakraborty2019cvd}
\Name{Chakraborty T., Lehmann F., Zhang J., Borgsdorf S., W{\"o}hrl N., Remfort R., Buck V., K{\"o}hler U. \and Suter D.} \REVIEW{Physical Review Materials}{3}{2019}{065205}.

\bibitem{xia2024design}
\Name{Xia S.-K., Lu W.-T., Zhao X.-T., Xue Y.-W., Xu Z.-B., Ge S.-Y., Wang Y., Yu L.-Y., Bian Y.-C., An S.-H. \etal} \REVIEW{Chinese Physics B}{}{2024}{}.

\bibitem{pogorzelski2024compact}
\Name{Pogorzelski J., Horsthemke L., Homrighausen J., Stiegek{\"o}tter D., Gregor M. \and Gl{\"o}sek{\"o}tter P.} \REVIEW{Sensors}{24}{2024}{743}.

\bibitem{stegemann2023modular}
\Name{Stegemann J., Peters M., Horsthemke L., Langels N., Gl{\"o}sek{\"o}tter P., Heusler S. \and Gregor M.} \REVIEW{European Journal of Physics}{44}{2023}{035402}.

\bibitem{misonou2020construction}
\Name{Misonou D., Sasaki K., Ishizu S., Monnai Y., Itoh K.~M. \and Abe E.} \REVIEW{AIP Advances}{10}{2020}{}.

\bibitem{zhao2023all}
\Name{Zhao M., Lin Q., Meng Q., Shan W., Zhu L., Chen Y., Liu T., Zhao L. \and Jiang Z.} \REVIEW{Nanomaterials}{13}{2023}{949}.

\bibitem{kim2019cmos}
\Name{Kim D., Ibrahim M.~I., Foy C., Trusheim M.~E., Han R. \and Englund D.~R.} \REVIEW{Nature Electronics}{2}{2019}{284}.

\bibitem{du2021high}
\Name{Du B., Huang K., Nie Y., Zhang Z., Xu R., Cui J. \and Li J.} \REVIEW{IEEE Sensors Journal}{21}{2021}{24665}.

\bibitem{qutools}
\Name{qutools} \Book{Quantum sensing by diamond magnetometer}.
\newline\url{https://qutools.com/qunv-deprecated/}

\bibitem{ciqtek}
\Name{ciqtek} \Book{Diamond quantum computer for education}.
\newline\url{https://www.ciqtekglobal.com/}

\bibitem{liu2018quantum}
\Name{Liu G.-Q. \and Pan X.-Y.} \REVIEW{Chinese Physics B}{27}{2018}{020304}.

\bibitem{balasubramanian2014nitrogen}
\Name{Balasubramanian G., Lazariev A., Arumugam S.~R. \and Duan D.-w.} \REVIEW{Current opinion in chemical biology}{20}{2014}{69}.

\bibitem{jelezko2001spectroscopy}
\Name{Jelezko F., Tietz C., Gruber A., Popa I., Nizovtsev A., Kilin S. \and Wrachtrup J.} \REVIEW{Single Molecules}{2}{2001}{255}.

\bibitem{nishimura2024investigations}
\Name{Nishimura S., Tsukamoto M., Sasaki K. \and Kobayashi K.} \REVIEW{arXiv preprint arXiv:2402.14422}{}{2024}{}.

\bibitem{scholten2021widefield}
\Name{Scholten S., Healey A., Robertson I., Abrahams G., Broadway D. \and Tetienne J.-P.} \REVIEW{Journal of Applied Physics}{130}{2021}{}.

\bibitem{taylor2008high}
\Name{Taylor J.~M., Cappellaro P., Childress L., Jiang L., Budker D., Hemmer P., Yacoby A., Walsworth R. \and Lukin M.} \REVIEW{Nature Physics}{4}{2008}{810}.

\bibitem{kost2015resolving}
\Name{Kost M., Cai J. \and Plenio M.~B.} \REVIEW{Scientific reports}{5}{2015}{11007}.

\bibitem{kudyshev2023machine}
\Name{Kudyshev Z.~A., Sychev D., Martin Z., Yesilyurt O., Bogdanov S.~I., Xu X., Chen P.-G., Kildishev A.~V., Boltasseva A. \and Shalaev V.~M.} \REVIEW{Nature communications}{14}{2023}{4828}.

\bibitem{oeckinghaus2014compact}
\Name{Oeckinghaus T., St{\"o}hr R., Kolesov R., Tisler J., Reinhard F. \and Wrachtrup J.} \REVIEW{Review of Scientific Instruments}{85}{2014}{}.

\bibitem{panadero2024photon}
\Name{Panadero I., Espin{\'o}s H., Tsunaki L., Volkova K., Tobalina A., Casanova J., Acedo P., Naydenov B., Puebla R. \and Torrontegui E.} \REVIEW{Physical Review Applied}{22}{2024}{014035}.

\bibitem{nizovtsev2001}
\Name{Nizovtsev A., Kilin S.~Y., Tietz C., Jelezko F. \and Wrachtrup J.} \REVIEW{Physica B: Condensed Matter}{308}{2001}{608}.

\bibitem{binder2017qudi}
\Name{Binder J.~M., Stark A., Tomek N., Scheuer J., Frank F., Jahnke K.~D., M{\"u}ller C., Schmitt S., Metsch M.~H., Unden T. \etal} \REVIEW{SoftwareX}{6}{2017}{85}.

\bibitem{zhang2020diamond}
\Name{Zhang J., Xu L., Bian G., Fan P., Li M., Liu W. \and Yuan H.} \REVIEW{arXiv preprint arXiv:2010.10231}{}{2020}{}.

\bibitem{irber2021robust}
\Name{Irber D.~M., Poggiali F., Kong F., Kieschnick M., L{\"u}hmann T., Kwiatkowski D., Meijer J., Du J., Shi F. \and Reinhard F.} \REVIEW{Nature Communications}{12}{2021}{532}.

\bibitem{wang2016coherence}
\Name{Wang J., Zhang W., Zhang J., You J., Li Y., Guo G., Feng F., Song X., Lou L., Zhu W. \etal} \REVIEW{Nanoscale}{8}{2016}{5780}.

\bibitem{tetienne2018spin}
\Name{Tetienne J.-P., De~Gille R., Broadway D., Teraji T., Lillie S., McCoey J., Dontschuk N., Hall L., Stacey A., Simpson D. \etal} \REVIEW{Physical Review B}{97}{2018}{085402}.

\bibitem{wang2012comparison}
\Name{Wang Z.-H., De~Lange G., Rist{\`e} D., Hanson R. \and Dobrovitski V.} \REVIEW{Physical Review B}{85}{2012}{155204}.

\bibitem{pham2012enhanced}
\Name{Pham L.~M., Bar-Gill N., Belthangady C., Le~Sage D., Cappellaro P., Lukin M.~D., Yacoby A. \and Walsworth R.~L.} \REVIEW{Physical Review B—Condensed Matter and Materials Physics}{86}{2012}{045214}.

\bibitem{storteboom2015lifetime}
\Name{Storteboom J., Dolan P., Castelletto S., Li X. \and Gu M.} \REVIEW{Optics express}{23}{2015}{11327}.

\bibitem{tisler2009fluorescence}
\Name{Tisler J., Balasubramanian G., Naydenov B., Kolesov R., Grotz B., Reuter R., Boudou J.-P., Curmi P.~A., Sennour M., Thorel A. \etal} \REVIEW{ACS nano}{3}{2009}{1959}.

\bibitem{hahn801950}
\Name{Hahn E.} \REVIEW{Rev}{80}{1950}{580}.

\end{thebibliography}
\bibliographystyle{eplbib}

\end{document}